\documentclass[twocolumn,showpacs,amsmath,amssymb,aps,
prd,nofootinbib]{revtex4}

\usepackage{graphicx}

\begin{document}

\title{Parametric Resonance and Cosmological Gravitational Waves}

\author{Paulo M. S\'a}

\email{pmsa@ualg.pt}

\affiliation{Centro Multidisciplinar de Astrof\'{\i}sica -- CENTRA and
Departamento de F\'{\i}sica, F.C.T., Universidade do Algarve, Campus Gambelas,
8005-139 Faro, Portugal}

\author{Alfredo B. Henriques}

\email{alfredo@fisica.ist.utl.pt}

\affiliation{Centro Multidisciplinar de Astrof\'{\i}sica -- CENTRA and
Departamento de F\'{\i}sica, Instituto Superior T\'ecnico, UTL, Av. Rovisco
Pais, 1049-001 Lisboa, Portugal}

\date{December 17, 2007}

\begin{abstract}
We investigate the production of gravitational waves due to quantum
fluctuations of the vacuum during the transition from the inflationary to the
radiation-dominated eras of the universe, assuming this transition to be
dominated by the phenomenon of parametric resonance. The energy spectrum of
the gravitational waves is calculated using the method of continuous
Bogoliubov coefficients, which avoids the problem of overproduction of
gravitons at large frequencies. We found, on the sole basis of the mechanism
of quantum fluctuations, that the resonance field leaves no explicit and
distinctive imprint on the gravitational-wave energy spectrum, apart an
overall upward or downward translation. Therefore, the main features in the
spectrum are due to the inflaton field, which leaves a characteristic imprint
at frequencies of the order of MHz/GHz.
\end{abstract}

 \maketitle

\section{Introduction}

Although gravitational waves have not yet been directly detected, those of
cosmological origin are at present the object of an important research effort,
as they will provide us with a unique telescope to the very early stages of
the evolution of the universe.

In this work we investigate the production of gravitational waves during the
transition from the inflationary period to the radiation-dominated era of the
universe. We assume this transition to be dominated by parametric resonance,
triggered by the coupling of the modes of an inhomogeneous bosonic field
$\chi$ to the homogeneous inflaton field $\phi$. We shall concentrate our
efforts on those gravitational waves created by quantum fluctuations of the
vacuum \cite{parker,grishchuk-starobinsky,allen}. This mechanism is
independent from the generation of gravitational waves directly sourced by
large inhomogeneities in matter distribution occurring during parametric
resonance \cite{khlebnikov-tkachev}. Although order of magnitude calculations
suggested that the last mechanism would dominate the production of
gravitational waves, during parametric resonance, we thought it would be of
interest to settle the issue with an accurate calculation based on the quantum
fluctuations of the vacuum.

Parametric resonance may play a fundamental role in the reheating
of the universe after the inflationary period
\cite{traschen-brandenberger}. Based on previous work on the
subject \cite{henriques-moorhouse}, we shall investigate a
simplified model of parametric resonance which, nevertheless,
keeps the most important features of the more complicated model.
For instance, this is the case when, as done in this paper, we
neglect the backreaction from the metric perturbations and also
perturbations of the inflaton field itself. This considerably
simplifies the calculations, without missing the most important
feature, namely, the sudden increase in the modes of the $\chi$
field, due to their coupling to the coherent oscillations of the
inflaton $\phi$.

We shall consider two different potentials for the inflaton field, namely, a
quadratic potential $V(\phi)=m_\phi^2 \phi^2/2$ and a quartic potential
$V(\phi)=\lambda \phi^4/4$. For the coupling between the inflaton field $\phi$
and the bosonic field $\chi$ we shall use $g^2\phi^2\chi^2/2$. The field
$\chi$ will be taken to have a mass term $V(\chi)=m_\chi^2\chi^2/2$. We
incorporate reheating through an elementary decay mechanism of the two scalar
fields into a relativistic radiation fluid characterized by a density
$\rho_{\mbox{\scriptsize rad}}$. The decay rates of the fields $\phi$ and
$\chi$ are determined by the decay constants $\Gamma_\phi$ and $\Gamma_\chi$,
respectively. Arguments have been given that the $\chi$ field is strongly
suppressed before the transition period begins \cite{l-l-m-w}. The enormous
increase of the scale factor during the inflationary era results in an
exceedingly small value for $\chi$ at the beginning of the transition. We are
then justified in treating the field $\chi$ as arising from its own quantum
fluctuations, without any primordial classical part.

Gravitational wave production will be calculated by the use of the Bogoliubov
coefficients as continuous functions of time, which obey differential
equations first introduced by Parker \cite{parker} (see also
Ref.~\cite{moorhouse-henriques-mendes} for a different derivation of these
differential equations). Besides being a theoretically correct method of
calculation of the graviton production, it has advantages over the frequently
used sudden transition approximation. Associated with the sudden transition
there is always an overproduction of gravitons of large frequencies, requiring
the introduction of an explicit cut-off for frequencies above the rate of
expansion of the universe. This problem is automatically solved by the use of
the continuous Bogoliubov coefficients \cite{mendes-henriques-moorhouse}.
During the inflationary period, the potential term of the inflaton field
dominates over the kinetic term, yielding a situation that closely resembles
the behavior of a fluid with an equation of state of the form $p=-\rho $, also
characterizing a cosmological constant. This, plus the fact that gravitational
waves produced during the inflationary period will be completely red-shifted,
simplifies the calculations, making it reasonable to begin our numerical
simulations at the end of the inflationary period, with simple initial
conditions for the Bogoliubov coefficients appropriate to a cosmological
constant.

This paper is organized as follows. In
Section~\ref{sect-gravwaves} we describe the method used to
calculate the gravitational-wave energy spectrum, present the
equations to determine the continuous Bogoliubov coefficients and
derive appropriate initial conditions for these equations. The
simplified model of parametric resonance, describing the
transition between the inflationary period and the
radiation-dominated era, as well as the equations describing the
evolution of the universe from the end of the transition period
till the present time, are introduced in
Section~\ref{sect-pararesonance}. Our numerical simulations, for
two different potentials of the inflaton field (quadratic and
quartic), are described in Section~\ref{sect-numerical}. Finally,
in Section~\ref{sect-conclusions}, we present our conclusions.

\section{Cosmological gravitational waves\label{sect-gravwaves}}

We assume a flat universe and write the perturbed metric, in
conformal time $\eta$, as
\begin{eqnarray}
ds^{2}=a^2(\eta ) \left\{ -d\eta^{2} + \left[ \delta_{ij} + h_{ij}
(\eta ,{\bf x}) \right] dx^i dx^j \right\},  \label{1}
\end{eqnarray}
where the tensor perturbations $h_{ij}$ are expanded in terms of plane waves
\begin{eqnarray}
h_{ij}(\eta,\textbf{x})=\sqrt{8\pi G}\sum\limits_{p=1}^{2}\int
\frac{d^3k}{(2\pi)^{3/2} a(\eta)\sqrt{2k}}\nonumber\\
\times \left[ a_p(\eta,{\bf k}) \varepsilon_{ij}(\textbf{k},p)
e^{i\textbf{k}\cdot\textbf{x}} \xi(\eta,k) + \mbox{H.c.}\right].
\end{eqnarray}
In the previous expression, $G$ is the gravitational constant, $p$ runs over
the two polarizations of the gravitational waves, $k=|\textbf{k}|=2\pi
a/\lambda =a\omega $ is the co-moving wave number, $a_p$ is the annihilation
operator, $\varepsilon_{ij}$ is the polarization tensor, and the mode function
$\xi$ obeys the equation of a parametric oscillator,
\begin{eqnarray}
\xi^{\prime\prime}+ \left( k^2-\frac{a^{\prime\prime}}{a} \right)
\xi = 0, \label{2}
\end{eqnarray}
where a prime denotes a derivative with respect to conformal time.
Under appropriate circumstances, we may have exponentially growing
solutions, with the gravitational field pumping energy into the
gravitational waves. Quantum mechanically this corresponds to
graviton production. We shall calculate the amount of gravitons
produced using Parker's method \cite{parker}.

In a expanding universe the ground-state and the operators change
with time,
\begin{eqnarray}
|0 \rangle_{\eta_i}\neq |0\rangle_{\eta},  \label{3}
\end{eqnarray}
meaning that the annihilation $a_p(\eta,{\bf k})$ and creation
$a^{\dag}_p(\eta,{\bf k})$ operators also change, such change being codified
in terms of time-fixed annihilation $A_p({\bf k})$ and creation
$A^{\dag}_p({\bf k})$ operators through the Bogoliubov transformation
\begin{eqnarray}
a_p(\eta,{\bf k}) = \alpha (\eta,k) A_p({\bf k}) + \beta^{*}(\eta,k)
A_p^{\dag}({\bf k}).  \label{4}
\end{eqnarray}
In the previous expression, the Bogoliubov coefficients $\alpha$ and $\beta$
obey the condition $|\alpha|^2-|\beta|^2=1$. Comparing any matter field
expansion at two different times, using Eq.~(\ref{4}) and the corresponding
one for $a_p^{\dag}(\eta,{\bf k})$, it can be shown that the coefficient
$\beta$ gives the number of gravitons created, $|\beta|^{2}=\langle N_{k}(\eta
)\rangle$.

To find the power spectrum, $P(\omega)$, we use the definition of the density
of states, $\omega^{2} d\omega /(2\pi^{2}c^{3})$, and the fact that each
graviton contributes with two polarizations, $2\hbar \omega$; then, from the
definition of the energy density in terms of the power spectrum, $dE=P(\omega
)d\omega$, we obtain
\begin{eqnarray}
P(\omega )=\frac{\hbar \omega ^{3}}{\pi ^{2}c^{3}} |\beta|^{2}. \label{7}
\end{eqnarray}

Taking into account that the gravitational-wave energy density is given by
$\rho_{\mbox{\scriptsize GW}} =\int P(\omega) d\omega$, we finally get the
dimensionless relative logarithmic energy spectrum of the gravitational waves
at the present time $\eta_0$:
\begin{eqnarray}
\Omega_{\mbox{\scriptsize GW}} (\omega,\eta_0) &=&
\frac{1}{\rho_{\mbox{\scriptsize crit}}(\eta_0)}
\frac{d\rho_{\mbox{\scriptsize GW}}}{d \ln\omega}(\eta_0) \nonumber \\
&=& \frac{8\hbar G}{3\pi c^5 H^2(\eta_0)} \omega^4 \beta^2(\eta_0), \label{8}
\end{eqnarray}
where $\rho _{\mbox{\scriptsize crit}}$ is the critical density of the
universe.

As we have just seen, in order to obtain the gravitational-wave
spectrum we need to compute the Bogoliubov coefficient $\beta$.
This is done by solving the set of differential equations
\cite{parker,moorhouse-henriques-mendes}
\begin{eqnarray}
\alpha^\prime &=& \frac{i}{2k}\left[ \alpha+\beta e^{2ik(\eta-\eta_i)} \right]
\frac{a^{\prime\prime}}{a},
\\
\beta^\prime &=& -\frac{i}{2k}\left[ \beta+\alpha e^{-2ik(\eta-\eta_i)}
\right] \frac{a^{\prime\prime}}{a},
\end{eqnarray}
which, upon the redefinition
\begin{eqnarray}
\alpha &=& \frac12 (X+Y) e^{ik(\eta -\eta_i)},\\
\beta &=& \frac12 (X-Y) e^{-ik(\eta -\eta_i)},  \label{5}
\end{eqnarray}
become
\begin{eqnarray}
X^\prime &=& -i k Y, \label{6a} \\
Y^\prime &=& -\frac{i}{k} \left( k^{2}-\frac{a^{\prime \prime }}{a} \right) X.
\label{6}
\end{eqnarray}

Because we are interested in graviton production during the
transition between the inflationary period and the
radiation-dominated era, the above set of differential equations
should be integrated from the end of the inflationary period right
up to the present time. This requires us to specify a model of
evolution of the universe for the entire period under
consideration in order to determine $a^{\prime \prime}/a$ (see
next section) and also to specify, at the end of the inflationary
period, the initial conditions for $X(\eta)$ and $Y(\eta)$.

Clearly, the set of differential equations (\ref{6a}) and
(\ref{6}) should be solved numerically. However, we can take
advantage of the fact that it admits an exact analytical solution
for the case of a de Sitter universe, in order to specify
appropriate initial conditions for $X(\eta)$ and $Y(\eta)$.
Indeed, during the inflationary period, the potential term of the
inflaton field dominates over the kinetic term, yielding a
situation that closely resembles the behavior of a cosmological
constant. Since, in this case, the scale factor is given by
$a(\eta)=[H(\eta_1-\eta)]^{-1}$, where $\eta<\eta_1$ ($\eta_1$ is
an arbitrary constant and $H$ is the Hubble parameter),
Eqs.~(\ref{6a}) and (\ref{6}) yield the solution
\begin{eqnarray}
X(\eta)&=& c_1 \left[ 1+\frac{i}{k(\eta_1-\eta)} \right] e^{ik(\eta_1-\eta)}
\nonumber \\
& & \hspace{-12mm} + \; c_2 \left[ 1-\frac{i}{k(\eta_1-\eta)} \right]
e^{-ik(\eta_1-\eta)}, \label{122a}
\\
Y(\eta)&=& c_1 \left[ 1 + \frac{i}{k(\eta_1-\eta)}
-\frac{1}{k^2(\eta_1-\eta)^2} \right] e^{ik(\eta_1-\eta)} \nonumber \\
& & \hspace{-12mm} + \; c_2 \left[ -1 + \frac{i}{k(\eta_1-\eta)}
+\frac{1}{k^2(\eta_1-\eta)^2} \right] e^{-ik(\eta_1-\eta)}. \label{122b}
\end{eqnarray}
The condition $|\alpha|^2-|\beta|^2=1$ imposes a constraint on the integration
constants $c_1$ and $c_2$, namely, $c_1^2-c_2^2=1$. Let us choose $c_1=1$ and
$c_2=0$. Then, at the end of the inflationary period,
\begin{eqnarray} \hspace{-3mm} X(\eta_{i})&=& \left(
1+\frac{i a(\eta_i)H}{k} \right) e^{ik/[a(\eta_i)H]}, \label{97}
\\
\hspace{-3mm} Y(\eta_{i})&=& \left( 1 + \frac{i a(\eta_i)H}{k}
-\frac{a^2(\eta_i)H^2}{k^2} \right) e^{ik/[a(\eta_i)H]}. \label{98}
\end{eqnarray}
These expressions will now be used as initial conditions.

Note that a slow-rolling inflaton field, as the one in our model
(see next section), does not mimic \emph{exactly} a cosmological
constant and, therefore, the Hubble parameter is not constant
during inflation, as assumed above. However, during the slow-roll
phase, the Hubble parameter decreases so slowly, that
Eqs.~(\ref{97}) and (\ref{98}), with $H=H(\eta_i)$, constitute a
very good approximation.

To finish this section, let us point out that measurements of the cosmic
microwave background radiation impose an upper limit on the energy spectrum
for angular frequencies corresponding to the present size of the horizon,
$h_0^2\Omega_{\mbox{\scriptsize GW}}(\omega _{\mbox{\scriptsize hor}},\eta_0)
\lesssim 7\times 10^{-11}$, where $h_0=H(\eta_0)/(100\mbox{ km
Mpc}^{-1}\mbox{s}^{-1})$ and $\omega _{\mbox{\scriptsize
hor}}=2\times10^{-17}h_0\mbox{ rad/s}$ \cite{allen,maggiore}. Timing
observations of milliseconds pulsars and Doppler tracking of the Cassini
spacecraft also provide bounds on the gravitational-wave energy spectrum,
respectively, $h_0^2\Omega_{\mbox{\scriptsize GW}}(\omega_{\mbox{\scriptsize
pul}},\eta_0) < 2.0\times 10^{-8}$ for $\omega_{\mbox{\scriptsize
pul}}=2.5\times10^{-8}\mbox{ rad/s}$ \cite{jenet-etal} and
$h_0^2\Omega_{\mbox{\scriptsize GW}}(\omega_{\mbox{\scriptsize Cas}},\eta_0) <
0.014$ for $\omega_{\mbox{\scriptsize Cas}}=7.5\times10^{-6}\mbox{ rad/s}$
\cite{armstrong}. For higher angular frequencies, of the order of a few
hundred rad/s, an upper limit is available from the Laser Interferometer
Gravitational Wave Observatory (LIGO), namely, $h_0^2\Omega_{\mbox{\scriptsize
GW}}(\omega,\eta_0) < 3.4\times10^{-5}$ \cite{LIGO}. Finally, an integral
bound can be derived from standard Big Bang nucleosynthesis,
$h_0^2\int_{\omega_n}^\infty \Omega_{\mbox{\scriptsize
GW}}(\omega,\eta_0)d\omega/\omega <5.6\times10^{-6}$, where
$\omega_n\thickapprox10^{-9}\mbox{ rad/s}$ \cite{allen,maggiore}.

\section{Parametric resonance\label{sect-pararesonance}}

Assuming that the transition between the inflationary and the radiation eras
is dominated by the phenomenon of parametric resonance, the Einstein equations
for the transition period are written in the form
\begin{eqnarray}
\left( \frac{a^{\prime}}{a}\right)^{2} &=& \frac{8\pi G}{3}a^{2}\rho
_{\mbox{\scriptsize tot}}, \label{10a}
\\
\frac{a^{\prime\prime}}{a} &=& \frac{4\pi G}{3} a^{2}(\rho _{\mbox{\scriptsize
tot}}-3p_{\mbox{\scriptsize tot}}), \label{10}
\end{eqnarray}
where the total energy density and pressure are given by
\begin{eqnarray}
&& \hspace{-9mm} \rho_{\mbox{\scriptsize tot}} = \frac{1}{a^{2}} \left[
\frac12 \phi^{\prime 2} + a^{2} V(\phi) + \frac12 \langle \chi^{\prime 2}
\rangle + \frac{1}{2}a^{2}m_{\chi }^{2} \langle \chi^{2}\rangle \right.
\nonumber \\
&& \hspace{7mm} + \left. \frac12 a^{2} g^{2} \langle \chi^{2} \rangle
\phi^{2}\right] + \rho _{\mbox{\scriptsize rad}}, \label{11} \\
&&
\hspace{-9mm} p_{\mbox{\scriptsize tot}} = \frac{1}{a^{2}} \left[ \frac12
\phi^{\prime 2} - a^{2} V(\phi) + \frac12 \langle \chi^{\prime 2} \rangle -
\frac{1}{2}a^{2}m_{\chi }^{2} \langle \chi^{2}\rangle \right.
\nonumber \\
&& \hspace{7mm} - \left. \frac12 a^{2} g^{2} \langle \chi^{2}
\rangle \phi^{2}\right] + \frac13\rho _{\mbox{\scriptsize rad}},
\label{12}
\end{eqnarray}
where $V(\phi)$ represents the quadratic and quartic potentials,
and $g^2\langle \chi^{2}\rangle\phi^2/2$ the coupling between the
two fields $\phi$ and $\chi$.

The equations for the inflaton and the relativistic radiation fluid are,
respectively,
\begin{eqnarray}
\phi^{\prime\prime} + 2\frac{a^{\prime}}{a}\phi^{\prime } + a^{2}
\partial_{\phi} V(\phi) + a^{2}g^{2}\langle \chi^{2}\rangle \phi =
-a \Gamma_{\phi}\phi^{\prime} \label{13}
\end{eqnarray}
and
\begin{eqnarray}
\rho_{\mbox{\scriptsize rad}}^{\prime} + 4\frac{a^{\prime}}{a}
\rho_{\mbox{\scriptsize rad}} = \frac{1}{a} \Gamma_{\phi} \phi^{\prime 2} +
\frac{1}{a} \Gamma_{\chi } \langle \chi^{\prime 2}\rangle.  \label{15}
\end{eqnarray}

Note that when terms involving $\chi^{2}$ appear in homogeneous equations,
like the ones above, the ensemble average
\begin{eqnarray}
\langle \chi ^{2} \rangle =\frac{1}{(2\pi)^{3}}\int d^{3}\kappa\chi
_{\kappa}\chi _{\kappa}^{*} \label{9}
\end{eqnarray}
should be used \cite{polarski-starobinsky}. The same applies to terms
involving $\chi^{\prime 2}$.

The non-homogeneous equation for the field $\chi$ is written in the
$\kappa$-component form, for the mode functions $\chi_\kappa$
\cite{henriques-moorhouse},
\begin{eqnarray}
\hspace{-2mm} \chi_{\kappa}^{\prime\prime} + 2\frac{a^{\prime}}{a}
\chi_{\kappa}^{\prime} + \left( \kappa^{2} + a^{2}m_{\chi}^{2} +
a^{2}g^{2}\phi^{2} \right) \chi_{\kappa} = -a
\Gamma_{\chi}\chi_{\kappa}^{\prime}.  \label{14}
\end{eqnarray}

Finally, the above set of equations is complemented with the equations
describing the evolution of the universe from the end of the transition period
till the present time, namely,
\begin{eqnarray}
\frac{a^{\prime\prime}}{a} &=&\frac{4\pi G}{3} a^2 \left[
\rho_{\mbox{\scriptsize mat}}(\eta_0) \left( \frac{a(\eta_0)}{a} \right)^3
\right.
\nonumber \\
&& + \left. (3w+1)\rho_{\mbox{\scriptsize de}} (\eta_0) \left(
\frac{a(\eta_0)}{a} \right)^{3(1-w)} \right], \label{101}
\\
\left( \frac{a^{\prime}}{a} \right)^2 &=& \frac{8\pi G}{3} a^2 \left[
\rho_{\mbox{\scriptsize rad}}(\eta_0) \left( \frac{a(\eta_0)}{a} \right)^4
\right.
\nonumber \\
&& \hspace{-20mm} \left.  +\, \rho_{\mbox{\scriptsize mat}}(\eta_0) \left(
\frac{a(\eta_0)}{a} \right)^3 +  \rho_{\mbox{\scriptsize de}}(\eta_0) \left(
\frac{a(\eta_0)}{a} \right)^{3(1-w)} \right], \label{102}
\end{eqnarray}
where for the dark energy we take $w=0.78$ and the density of
radiation, matter and dark energy at the present time are,
respectively, $\rho_{\mbox{\scriptsize rad}}(\eta_0)=4.6\times
10^{-31} \mbox{ kg/m}^3$, $\rho_{\mbox{\scriptsize
mat}}(\eta_0)=2.6\times10^{-27} \mbox{ kg/m}^3$ and
$\rho_{\mbox{\scriptsize de}}(\eta_0)=6.9\times10^{-27} \mbox{
kg/m}^3$.

The above set of equations (\ref{10a})--(\ref{102}), together with
Eqs.~(\ref{6a}) and (\ref{6}) for the Bogoliubov coefficients, are solved
numerically in order to obtain the energy spectrum of the gravitational waves.

\section{Numerical simulations\label{sect-numerical}}

Our numerical simulations begin at the end of the inflationary period and
continue up to the present epoch. We use a Runge-Kutte method with variable
step to solve the system of differential equations (\ref{10})--(\ref{101}).
Equations (\ref{10a}) and (\ref{102}) are used to check the accuracy of the
numerical solution. The integral in Eq.~(\ref{9}) is computed using the
Simpson rule, with a cut-off at $\kappa=2\pi a H$ and a interval of
integration divided in 20 segments of equal length $h=(\pi/10)a^{\prime}/a$.
Having determined the time evolution of $a^{\prime\prime}/a$, we then solve
numerically Eqs.~(\ref{6a}) and (\ref{6}) (again with a Runge-Kutte method
with variable step) for different values of $\omega=k/a$, compute $\beta^2$
with Eq.~(\ref{5}) and, finally, obtain $\Omega_{\mbox{\scriptsize
GW}}(\omega,\eta_0)$ from Eq.~(\ref{8}).

The energy spectrum of the gravitational waves, $\Omega_{\mbox{\scriptsize
GW}}(\omega,\eta_0)$, is computed for values of $\omega$ in the interval
$\omega_{\mbox{\scriptsize min}}\leq \omega\leq \omega_{\mbox{\scriptsize
max}}$. The lower limit corresponds to a gravitational wave with a wavelength
equal, today, to the Hubble distance: $\omega_{\mbox{\scriptsize min}}=2\pi
c/d_{\mbox{\scriptsize Hubble}}(\eta_0)\approx 2\pi
H(\eta_0)=1.4\times10^{-17} \mbox{ rad/s}$, where the Hubble parameter at the
present time was taken to be $H(\eta_0)=71\mbox{ km s}^{-1}\mbox{Mpc}^{-1}$.
The upper limit corresponds to a gravitational wave which, at the beginning of
the transition period between the inflationary and radiation-dominated eras,
had a wavelength equal to the Hubble distance at that time; because of
red-shifting, the angular frequency of this gravitational wave is, today,
$\omega_{\mbox{\scriptsize max}}\approx 2\pi H(\eta_i)[a(\eta_i)/a(\eta_0)]$,
where $\eta_i$ is the conformal time at the beginning of the transition
period. Within the model we are considering, the maximum angular frequency is
about $(10^9-10^{10})\mbox{ rad/s}$.

There are two more characteristic values of $\omega$ in the gravitational-wave
energy spectrum, corresponding to gravitational waves which had a wavelength
equal to the Hubble distance at the moments
$\eta_{\phi\rightarrow\mbox{\scriptsize rad}}$ and $\eta_{\mbox{\scriptsize
rad}\rightarrow\mbox{\scriptsize mat}}$, these being, respectively, the time
when the energy density of the inflaton becomes equal to the energy density of
radiation (marking the end of the transition period between the inflationary
and radiation-dominated eras) and the time when the energy density of
radiation becomes equal to the energy density of matter (corresponding to the
transition from a radiation-dominated to a matter-dominated universe). These
angular frequencies are, today, $\omega_{\phi\rightarrow\mbox{\scriptsize
rad}} \approx 10^7 \mbox{ rad/s}$ and $\omega_{\mbox{\scriptsize rad}
\rightarrow \mbox{\scriptsize mat}} \approx 10^{-15} \mbox{ rad/s}$.

The gravitational-wave energy spectrum is then naturally divided in three
regions. The first, from $\omega_{\mbox{\scriptsize min}}$ to
$\omega_{\mbox{\scriptsize rad} \rightarrow \mbox{\scriptsize mat}}$, marked
by a sudden decrease of $\Omega_{\mbox{\scriptsize GW}}$, the second, from
$\omega_{\mbox{\scriptsize rad} \rightarrow \mbox{\scriptsize mat}}$ to
$\omega_{\phi \rightarrow \mbox{\scriptsize rad}}$, where
$\Omega_{\mbox{\scriptsize GW}}$ is constant (corresponding to a
radiation-dominated universe, in which there is no production of gravitational
waves), and the third, from $\omega_{\phi \rightarrow \mbox{\scriptsize rad}}$
to $\omega_{\mbox{\scriptsize max}}$, where $\Omega_{\mbox{\scriptsize GW}}$
has some complex structure (peaks and valleys), which depends on the details
of the transition from the inflationary period to the radiation-dominated era.

Our investigation is carried out for two different potentials of the inflaton
field $\phi$, namely, $V(\phi)=m_{\phi}^{2}\phi^{2}/2$ and $V(\phi)=\lambda
\phi^{4}/4$. The values of the parameters $m_{\phi}$ and $\lambda$ are chosen
such that the bounds from the temperature anisotropy of the cosmic microwave
background radiation (CMBR) and from large-scale structure (LSS) are satisfied
\cite{smith-kam-cooray},
\begin{eqnarray}
m_\phi=\frac{\sqrt{3\pi P_s(k_c)}}{2N(k_c)+1} m_{\mbox{\scriptsize pl}}
\end{eqnarray}
and
\begin{eqnarray}
\lambda=\frac{3\pi^2 P_s(k_c)}{2[N(k_c)+1]^3}.
\end{eqnarray}
In the previous expressions, $P_s(k)$ is the power spectrum for density
perturbations and $N(k)$ is the number of $e$-foldings of expansion between
the time when a co-moving distance scale labelled by $k$ exited the horizon
during inflation and the end of inflation, both $P_s(k)$ and $N(k)$ being
evaluated at the CMBR/LSS scale $k_c=0.05\mbox{ Mpc}^{-1}$. For $47\leq
N(k_c)\leq 62$ and $P_s(k_c)=(2.45\pm0.23)\times10^{-9}$
\cite{smith-kam-cooray}, we obtain the constraints $1.2\times10^{-6} \leq
m_\phi/m_{\mbox{\scriptsize pl}} \leq 1.7\times10^{-6}$ and $1.3 \times
10^{-13} \leq \lambda \leq 3.6\times 10^{-13}$.

In the slow-roll approximation, inflation ends when the parameter
$\epsilon\equiv[m_{\mbox{\scriptsize pl}}^2/(16\pi)] (\partial_\phi V/V)^2
\approx 1$, implying that $\phi(\eta_i) \approx m_{\mbox{\scriptsize
pl}}/(2\sqrt{\pi})$ for the case $V(\phi)=m_\phi^2\phi^2/2$ and $\phi(\eta_i)
\approx m_{\mbox{\scriptsize pl}}/\sqrt{\pi}$ for the case
$V(\phi)=\lambda\phi^4/4$. The initial value of $\phi^\prime$ is chosen such
that $\phi^{\prime2}(\eta_i)/[2a^2(\eta_i)]\lesssim V[\phi(\eta_i)]$. We also
choose $\chi_{\kappa}(\eta_i)$ and $\chi^\prime_{\kappa}(\eta_i)$, as well as
$m_\chi$ and $g$, such that $\rho_{\mbox{\scriptsize tot}}$ [see
Eq.~(\ref{11})] is dominated by the potential and kinetic terms of the
inflaton field. Since any pre-existing radiation fluid is dilute during
inflation, we choose $\rho_{\mbox{\scriptsize rad}}(\eta_i)=0$. Finally, the
decay constants $\Gamma_\phi$ and $\Gamma_\chi$ are chosen to be small enough
to allow a significant growth of the resonant field $\chi$ before the
evolution of the universe becomes dominated by radiation.

\subsection{Case $V(\phi)=\frac12m_\phi^2\phi^2$}

The masses of the fields $\phi$ and $\chi$, the coupling constant $g$ and the
decay constants $\Gamma_\phi$ and $\Gamma_\chi$ are chosen to take the
following values: $m_\phi=1.5\times10^{-6} \, m_{\mbox{\scriptsize pl}}$,
$m_\chi=10^{-7} \, m_{\mbox{\scriptsize pl}}$, $0\leq g\leq 10^{-2}$,
$\Gamma_\phi=\Gamma_\chi=5\times10^{-10} \, m_{\mbox{\scriptsize pl}}$. As
initial conditions, at the beginning of the transition period, we choose
$a(\eta_i)=1$, $\phi(\eta_i)=0.3 \, m_{\mbox{\scriptsize pl}}$,
$\phi^\prime(\eta_i)=-4.5\times10^{-7} \, m_{\mbox{\scriptsize pl}}^2$,
$\rho(\eta_i)=0$, $\chi_{\kappa}(\eta_i)=10^{-25}$ and
$\chi^\prime_{\kappa}(\eta_i)=0$; the constraint equation (\ref{10a}) gives
then $a^\prime(\eta_i)\approx 1.30\times10^{-6} \, m_{\mbox{\scriptsize pl}}$.

The relevant parameter to consider is the resonance para\-me\-ter, defined as
$q\equiv g^{2}\phi^2/(4m_{\phi,\mbox{\scriptsize eff}}^{2})$, with the
effective mass of the inflaton given by $m_{\phi,\mbox{\scriptsize eff}}^2 =
m_{\phi}^{2} + g^{2} \langle\chi^2\rangle$. In order to achieve a significant
resonance we need values of $q>1$, the so-called broad resonance regime.

Our numerical simulations show that the energy density of the field $\chi$,
$\rho_\chi=\langle\chi^{\prime2}\rangle/(2a^2) + m_\chi^2
\langle\chi^2\rangle/2$, increases by many orders of magnitude in a short time
interval, reaching values comparable to the values of the energy density of
the inflaton field, $\rho_\phi=\phi^{\prime2}/(2a^2) + m_\phi^2 \phi^2/2$. In
Fig.~1 we can see that jump in the value of $\rho_{\chi}$. However, when
switching off the resonant field $\chi$, the most relevant change occurring in
the spectrum $\Omega_{\mbox{\scriptsize GW}}$ is an overall translation
upwards or downwards (see Fig.~2).

\begin{figure}
\centering
\includegraphics[width=8.6cm]{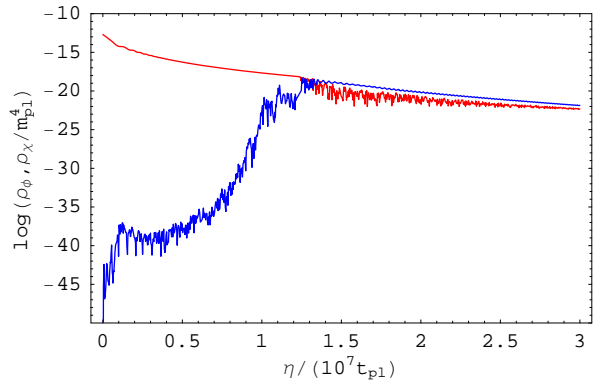}
\caption{(Color online) The jump in the value of $\rho_\chi$.
After a short time interval, the energy density of the field
$\chi$ (lower/blue curve) becomes comparable to the energy density
of the inflaton (upper/red curve). Both curves refer to case with
$g=6.5\times10^{-3}$.} \label{fig-1}
\end{figure}

\begin{figure}
\centering
\includegraphics[width=8.6cm]{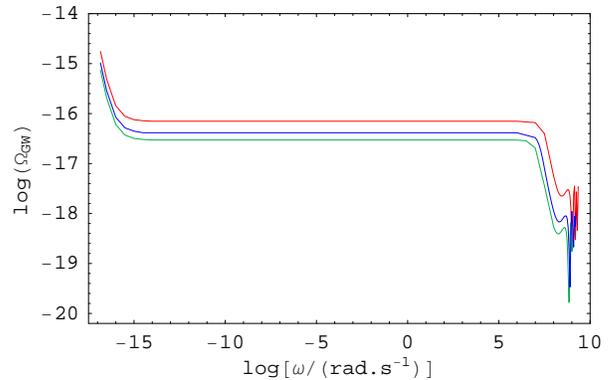}
\caption{(Color online) Gravitational-wave spectra for the model
with $V(\phi)=m_\phi^2 \phi^2/2$. The effect of switching on the
resonant field $\chi$ is an overall translation upwards or
downwards. Upper/red curve refers to the case with
$g=6.5\times10^{-3}$, middle/blue curve to $g=0$ (field $\chi$
switched off) and lower/green curve to $g=10^{-3}$.} \label{fig-2}
\end{figure}

No explicit and distinctive signal left by the resonance can be read from
$\Omega_{\mbox{\scriptsize GW}}$. At first we assumed that this was due to the
fact that $\rho_{\chi}$ did not dominate over $\rho_{\phi}$ and
$\rho_{\mbox{\scriptsize rad}}$ for a sufficiently long period of time, and
that, by increasing $g$ (and, therefore, the resonant parameter $q$), this
could be solved. This does not happen and, from Figs.~3 and 4, we may
understand why: when $\langle\chi^{2}\rangle$ begins to increase, the same
happens to the effective mass of the inflaton, due to the presence of the term
$g^2\langle\chi^{2}\rangle$; as a consequence, the value of $q$ decreases and
the resonance is killed.

\begin{figure}
\centering
\includegraphics[width=8.6cm]{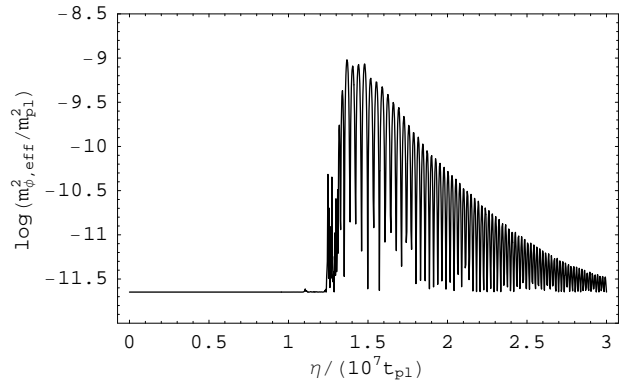}
\caption{The change in time of the effective mass of the inflaton. Its value
increases sharply, when the energy density of the resonant field $\chi$
becomes comparable to the energy density of the field $\phi$.} \label{fig-3}
\end{figure}

\begin{figure}
\centering
\includegraphics[width=8.6cm]{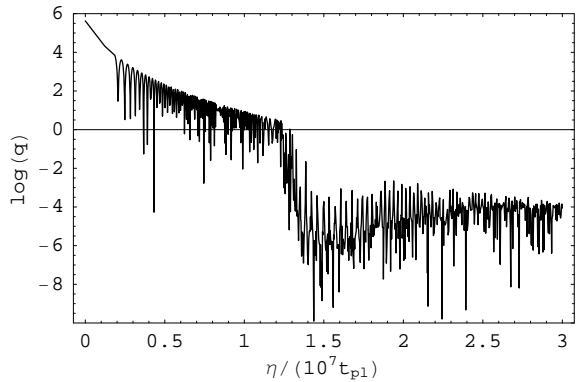}
\caption{The change in the resonance parameter $q$. Because of the sharp
increase of the effective mass of the inflaton, the resonance parameter
suddenly decreases to values much smaller than the unity, $q\ll 1$; as a
consequence, the resonance regime ends.} \label{fig-4}
\end{figure}

As expected, the transition between the inflationary period and
the radiation-dominated era leaves its imprint in the
gravitational-wave spectrum $\Omega_{\mbox{\scriptsize GW}}$ at
frequencies of the order of MHz/GHz. However, this imprint is due
to the form of the potential of the inflaton field and to the
values of the parameters $m_\phi$ and $\Gamma_\phi$; the presence
of the resonant field $\chi$ only induces an overall upward or
downward translation of the gravitational-wave spectrum. This
upward or downward translation depends on the values of the mass
of the resonant field $m_\chi$, the coupling constant $g$ and the
decay constant $\Gamma_\chi$. In Fig.~5 we compare the energy
spectrum in the presence of the resonant field $\chi$ (for $0<
g\leq10^{-2}$) with the energy spectrum in the absence of this
field. The comparison is made at the flat part of the spectrum,
namely, at $\omega=10^2\mbox{ rad/s}$. As clearly seen in Fig.~5,
the spectrum tends to move downwards for small values of $g$,
while for large values of $g$ it tends to move upwards.

\begin{figure}
\centering
\includegraphics[width=8.6cm]{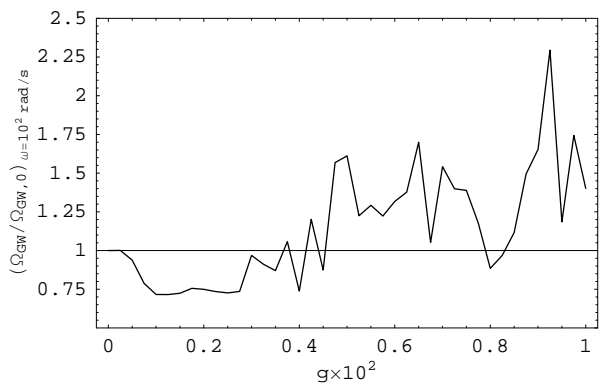}
\caption{A manifestation of the bands of resonance in the model with $V(\phi
)=m_\phi^2 \phi^2/2$. $\Omega_{\mbox{\tiny GW},0}$ denotes the energy spectrum
of the gravitational waves in the absence of the resonant field $\chi$. The
ratio $\Omega_{\mbox{\tiny GW}}/\Omega_{\mbox{\tiny GW},0}$ is taken at the
flat part of the spectrum, namely, at $\omega=10^2\mbox{ rad/s}$. For small
values of $g$ the spectrum tends to move downwards relatively to the case in
which the field $\chi$ is switched off, while for large values of $g$ we have
the opposite effect.} \label{fig-5}
\end{figure}

The results presented above indicate that, as far as the creation of gravitons
due to quantum fluctuations of the vacuum is concerned, the main features in
the spectrum $\Omega_{\mbox{\scriptsize GW}}(\omega,\eta_0)$ are due to the
inflaton field. This is different from what happens when we consider
gravitational waves directly produced by the motion of matter. In that case,
the time-dependent inhomogeneities in the distribution of matter, produced by
the rapid growth of the resonant field $\chi$, leave a much stronger imprint
in the energy spectrum of the gravitational waves $\Omega_{\mbox{\scriptsize
GW}}$ at frequencies of the order of MHz/GHz \cite{khlebnikov-tkachev}.

\subsection{Case $V(\phi )=\frac{1}{4}\lambda \phi ^{4}$}

The parameters are chosen to take the following values:
$\lambda=3.5\times10^{-13}$, $m_\chi=10^{-9} \, m_{\mbox{\scriptsize pl}}$,
$0\leq g\leq 2.8\times10^{-5}$, $\Gamma_\phi=\Gamma_\chi=3\times10^{-12}$. As
initial conditions, at the beginning of the transition period, we choose
$a(\eta_i)=1$, $\phi(\eta_i)=0.6 \, m_{\mbox{\scriptsize pl}}$,
$\phi^\prime(\eta_i)=-1.5\times10^{-7} \, m_{\mbox{\scriptsize pl}}^2$,
$\rho(\eta_i)=0$, $\chi_{\kappa}(\eta_i)=10^{-25}$ and
$\chi^\prime_{\kappa}(\eta_i)=0$; the constraint equation (\ref{10a}) gives
then $a^\prime(\eta_i)\approx 4.35\times10^{-7} \, m_{\mbox{\scriptsize pl}}$.

In Fig.~6 we show two spectra, one with and the other without the field
$\chi$. The same conclusions can be drawn as presented in the previous
subsection: no obvious imprint of the resonance of the field $\chi$ in the
energy spectrum of the gravitational waves, apart from the overall
displacement.

\begin{figure}
\centering
\includegraphics[width=8.6cm]{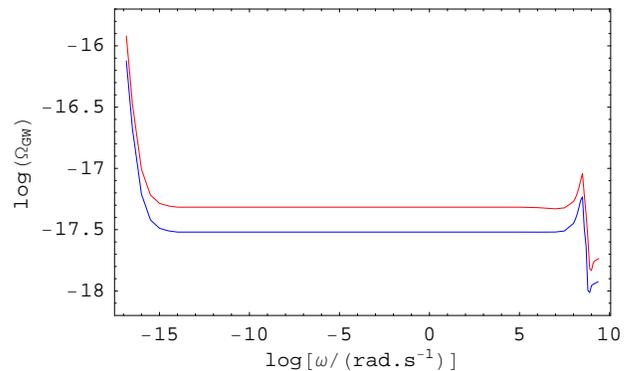}
\caption{(Color online) Gravitational-wave spectra for the model
with  $V(\phi )=\lambda \phi^4/4$. Upper/red curve refers to the
case with $g=10^{-5}$ and lower/blue curve to $g=0$ (field $\chi$
switched off).} \label{fig-6}
\end{figure}

In Fig.~7 we show the bands of resonance which, in this case, are much better
marked due to the fact that $q$ is now the ratio between constant parameters,
$q\equiv g^2/(4\lambda)$.

\begin{figure}
\centering
\includegraphics[width=8.6cm]{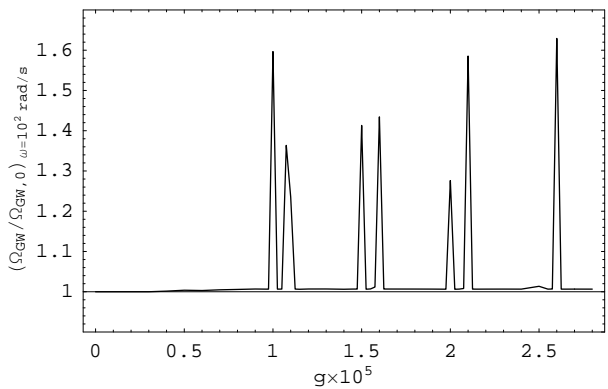}
\caption{The structure of bands of resonance in the model with
$V(\phi)=\lambda \phi^4/4$. $\Omega_{\mbox{\tiny GW},0}$ denotes the energy
spectrum of the gravitational waves in the absence of the resonant field
$\chi$. The ratio $\Omega_{\mbox{\tiny GW}}/\Omega_{\mbox{\tiny GW},0}$ is
taken at the flat part of the spectrum, namely, at $\omega=10^2\mbox{
rad/s}$.} \label{fig-7}
\end{figure}

If we compare Figs.~2 and 6, something interesting can be observed. While in
Fig.~6 we see the oscillations in the MHz/GHz region to be around the value
defined by the flat part of the spectrum, in Fig.~2 there is first a
pronounced decrease, followed then by the usual oscillations. This can be
qualitatively understood as follows. Given an homogeneous scalar field with a
potential $V(\phi) \propto \phi^n$, oscillating rapidly relatively to the
expansion rate of the universe, for $n=2$ the energy density of the
scalar-field oscillations behaves like non-relativistic matter (with $p=0$),
while for $n=4$ it behaves like relativistic matter with equation of state
$p=\rho/3$ \cite{turner}. Now, it is known \cite{henriques} that the first
case leads to a decrease in the energy density of the gravitational waves,
while the other tends to keep the energy density at the level of the
horizontal part of spectrum.

\section{Conclusions\label{sect-conclusions}}

In this work we have investigated the production of
gravi\-tational waves due to quantum fluctuations of the vacuum
during the transition between the inflationary period and the
radiation-dominated era of the universe, assuming that the
phenomenon of parametric resonance dominated this transition. The
energy spectrum of the gravitational waves was calculated using
the \textit{continuous} Bogoliubov coefficients. This approach
automatically solves the problem of overproduction of gravitons at
high frequencies which afflicts the sudden-transition
approximation.

We have found that the resonance field $\chi$ leaves no explicit and
distinctive imprint on the gravitational-wave energy spectrum, apart an
overall upward or downward translation. The amplitude of this translation,
which depends on the values of the mass of the resonant field $m_\chi$, the
coupling constant $g$ and the decay constant $\Gamma_\chi$, is quite modest.
Consequently, as far as the production of gravitons due to quantum
fluctuations of the vacuum is concerned, the main features in the spectrum are
due to the inflaton field, which leaves a characteristic imprint at
frequencies of the order of MHz/GHz. For the models of chaotic inflation we
have considered (quadratic and quartic potentials), the relative energy
density of the gravitational waves in this frequency range is quite small,
$\Omega_{\mbox{\scriptsize GW}}\lesssim 10^{-17}$. Therefore, our calculations
show that, by far, the main contribution to the production of gravitons in the
MHz/GHz frequency range, for this type of models, does come from the direct
coupling of the anisotropic stress tensor, describing the motion of particles
during preheating, as investigated in Ref.~\cite{khlebnikov-tkachev}.

The MHz/GHz region of the spectrum lies far beyond the range of frequencies
accessible to ground- or space-based interferometer detectors, as well as to
resonant bars and spheres (spanning roughly from $10^{-4}$ to $10^4$ Hz). In
recent years, considerable effort has been made in order to develop
high-frequency gravitational wave detectors \cite{cruise-ingley}. In a near
future, these detectors may reach sensitivities high enough to explore this
region of the gravitational-wave spectrum.

Below frequencies of the order of MHz, the influence of the transition regime
between the inflationary period and the radiation-dominated era should not be
felt anymore and, unless strong anisotropies develop later on, the main source
of gravitational waves will then be the quantum fluctuations of the vacuum.
However, the level of the flat part of the spectrum is mainly fixed by the
parameters defining the kinetic and potential terms of the inflaton field,
which are known to obey severe constraints from the observations. What our
calculation indicate is that the two popular models of inflation we have used
will not be able to generate enough gravitational waves to be seen by the
interferometer detectors LIGO, Virgo and LISA or in the cosmic microwave
background radiation.

\begin{acknowledgments}
The authors would like to thank Prof. James Hough for calling their attention
to the possibilities concerning the detection of gravitational waves in the
MHz/GHz region. PMS also thanks R. Potting for interesting discussions.
\end{acknowledgments}

\end{document}